\documentclass[aps,prl,twocolumn,superscriptaddress,showpacs,nobalancelastpage]{revtex4}

\usepackage{graphicx}
\usepackage{amsmath}

\begin{document}

\title{Detection of Neutrinos from Supernovae in Nearby Galaxies}

\author{Shin'ichiro Ando}
\affiliation{Department of Physics, Ohio State University, Columbus, Ohio
43210, USA}
\affiliation{Department of Physics, School of Science, University of
Tokyo, Tokyo 113-0033, Japan}

\author{John F. Beacom}
\affiliation{Department of Physics, Ohio State University, Columbus, Ohio
43210, USA}
\affiliation{Department of Astronomy, Ohio State University, Columbus, Ohio
43210, USA}

\author{Hasan Y{\"u}ksel}
\affiliation{Department of Physics, Ohio State University, Columbus, Ohio
43210, USA}
\affiliation{Department of Physics, University of Wisconsin, Madison,
Wisconsin 53706, USA}

\date{15 March 2005; revised 26 July 2005}

\begin{abstract}
While existing detectors would see a burst of many neutrinos
from a Milky Way supernova, the supernova rate is only
a few per century.  As an alternative, we propose the detection of
$\sim 1$ neutrino per supernova from galaxies within 10 Mpc, in which
there were at least 9 core-collapse supernovae since 2002.
With a future 1-Mton scale detector, this could be a
faster method for measuring the
supernova neutrino spectrum, which is essential for calibrating
numerical models and predicting the redshifted diffuse spectrum from
distant supernovae.  It would also allow a $\gtrsim 10^4$ times more
precise trigger time than optical data alone for high-energy neutrinos
and gravitational waves.
\end{abstract}



 \maketitle


%
One of the unsolved problems of astrophysics is how
core-collapse supernovae explode.  Nuclear fusion reactions in the
core of a massive star produce progressively heavier elements until a
Chandrasekhar mass of iron is formed, and electron degeneracy pressure
cannot support the core under the weight of the stellar envelope.  The
core collapses until it reaches nuclear densities and neutrino
emission begins; then an outgoing bounce shock should form, unbinding
the envelope and producing the optical supernova.  While
successful in nature, in most numerical supernova models, the shock stalls,
so that the fate of the entire star is to produce a black hole (after
substantial neutrino emission), but no optical supernova.

Since the gravitational energy release transferred to neutrinos, about
$3 \times 10^{53}$ erg, is $\sim 100$ times greater than the required
kinetic energy for the explosion, it is thought that neutrino emission
and interactions are a key diagnostic or ingredient of success.
However, not enough is directly known about the total energies and
temperatures of the neutrino flavors.  The $\simeq 20$ events from SN
1987A were only crudely consistent with expectations for
$\bar{\nu}_e$, and gave very little information on the other flavors~\cite{nu87A}.
It is thus essential to collect more supernova neutrino events.  A
Milky Way supernova would allow detailed measurements, but the
supernova rate is only a few per century.  If Super-Kamiokande were
loaded with GdCl$_3$~\cite{GADZOOKS}, the diffuse supernova neutrino
background (DSNB)~\cite{Ando2002,DSNB,Concordance} could be cleanly
detected, probing the supernova neutrino spectrum, but convolved with
the rapidly evolving star formation rate~\cite{GALEX} up to redshift $z \simeq 1$.

\begin{figure}[t]
\includegraphics[width=3.25in,clip=true]{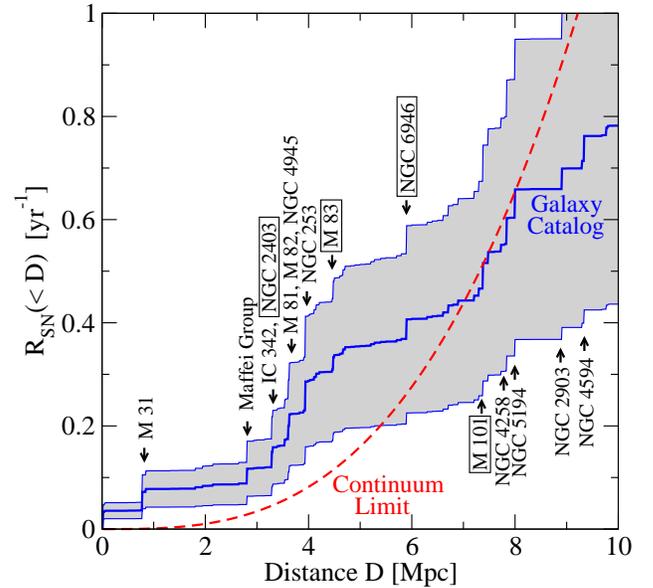}
\caption{Cumulative calculated core-collapse supernova rate versus
distance.  The dashed line is the continuum limit using the GALEX $z =
0$ star formation rate~\cite{GALEX}.  For our particular local volume,
and its fortuitous enhancement, we use a galaxy
catalog~\cite{Catalog}; the stepped line is based on star formation
rates for individual galaxies, and the band is the uncertainty.  Some
major galaxies are indicated, and those in boxes have especially high
optical supernova rates (see Table~\ref{table:recentsn}).
\label{fig:snrate}}
\end{figure}

We propose an intermediate regime, in which the number of events per
supernova is $\sim 1$, instead of $\gg 1$ (Milky Way) or $\ll 1$
(DSNB), motivated by the serious consideration of 1-Mton
scale water-\v{C}erenkov detectors in Japan (Hyper-Kamiokande~\cite{HK}),
the United States (UNO~\cite{UNO}), and Europe
(MEMPHYS~\cite{MEMPHYS}).  These detectors, which may operate for
decades, are intended for proton decay and long-baseline accelerator
neutrino oscillation studies, but could also detect neutrinos from
Milky Way supernovae, a point which has attracted much
interest~\cite{SNMton}.  The distance range of a 1-Mton detector is about
10 Mpc, within which the calculated supernova rate is about one per year, as shown
in Fig.~\ref{fig:snrate}.  Since the number of events per supernova is small,
background rejection requires a coincidence of at least two neutrinos or one
neutrino and an optical (or other waveband) supernova.

\begin{figure}[t]
\includegraphics[width=3.25in,clip=true]{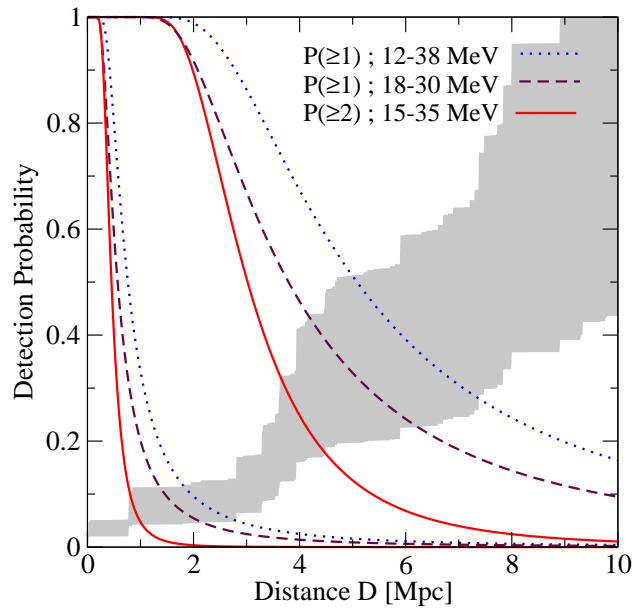}
\caption{Probability of detecting at least one (dotted and dashed
curves) or at least two (solid curves) supernova neutrinos versus
distance.  Background considerations restrict the useful energy
intervals, labeled here, and explained in the text.  The upper set of
curves is for a 1-Mton detector, and the lower set for the 22.5 kton
Super-Kamiokande; for another detector size, scale the distance
to compensate.  The shaded band indicates the calculated cumulative
supernova rate.  The probabilities for single background events are
$\simeq 0.4$ (1-Mton) and $\simeq 0.02$ (Super-Kamiokande),
independent of distance.
\label{fig:prob}}
\end{figure}

{\bf Supernova Neutrino Detection.---}%
For a Milky Way supernova at 10 kpc, the expected number of events in
Super-Kamiokande (22.5 kton) is $\sim 10^4$, corresponding to
1 event at 1 Mpc, 0.1 events at 3 Mpc, and so on.  For an expected number
of events $\mu$, the Poisson probability to detect $n$ events is
$P_n = \mu^n e^{-\mu} / n!$; for small $\mu$, we scale $P_1 \simeq \mu$
by the number of supernovae.    As shown in Fig.~\ref{fig:prob},
for each supernova within, say 4 Mpc, the chance of
detecting a single neutrino (or a background event; see below) in
Super-Kamiokande is $\sim 3\%$.  While small, this should motivate a
careful analysis of their data.

To make this technique more efficient, detectors larger than
Super-Kamiokande are needed.
We consider a similar detector with a 1-Mton
fiducial volume, which is somewhat larger than the proposed detectors,
but if two are built, the combined mass could exceed 1 Mton.  In
Fig.~\ref{fig:prob}, we show the detection probabilities for at least
one or two events from a single supernova versus distance, along with
the calculated supernova rate, which coincidentally also varies from 0
to 1.  For a 1-Mton detector, both the detection probability per
supernova and the relevant supernova rate are quite favorable, so that
the supernova neutrino spectrum could be constructed,
slowly but (almost) steadily.  Additionally, the detection of even a
single neutrino could fix the start time of the supernova to $\sim 10$
seconds instead of $\sim 1$ day, greatly reducing backgrounds for
observing prompt gravitational wave or high energy neutrino emission.
Calculations of the nearby supernova rate and background rejection
capabilities are needed, and we turn to these next.

{\bf Nearby Supernova Rate.---}%
The supernova rate within a typical sphere of radius 10 Mpc can be
calculated using the $z = 0$ limit of the measured star formation
rate, for which we use the latest dust-corrected measurements from
GALEX~\cite{GALEX} (other recent measurements are in agreement).  We
convert this to a core-collapse supernova rate using the stellar
initial mass function to calculate the fraction of stars above $8
M_\odot$; the result is in good agreement with the measured
core-collapse supernova rate versus redshift, as shown in
Ref.~\cite{Concordance}.  Our ``Continuum Limit" result is shown in
Fig.~\ref{fig:snrate}.  Since galaxies are clustered and have varying
supernova rates, our local volume may differ from a typical volume.
It does, and in fact the nearby supernova rate is higher than typical.
We used the recent catalog of Ref.~\cite{Catalog} to obtain
galaxy distances, morphological types, and optical luminosities, and
then the conversion factors of Ref.~\cite{Cappellaro} to calculate the
supernova rate for each galaxy; our ``Galaxy Catalog" result is also
shown in Fig.~\ref{fig:snrate} (see also Ref.~\cite{Arnaud}).  Most of
the uncertainty comes from the conversion between galaxy properties
and supernova rate, and could be substantially reduced by direct
measurements of the star formation rates for these specific galaxies.

The calculated core-collapse supernova rate
within 10 Mpc is about one per year; this arises both from many
galaxies similar to the Milky Way, as well as several indicated
galaxies with higher rates (Table~\ref{table:recentsn} lists galaxies
with especially high historical supernova rates). 
Our calculations are based on {\it star formation} rates, which should predict
{\it supernova} rates (type Ia supernovae, which do not have
substantial neutrino emission, are only about 15\% of supernovae).
If there were bursts of star formation on timescales less
than the lifetimes of massive stars, these results could differ.
Quite recently, due to a rise in the quantity and quality of supernova
searches, the number of discovered supernovae has increased very
dramatically~\cite{sn-list-cbat}, strongly suggesting that the
calculated and historical supernova rates are significant underestimates.  
Since 2002, there were at least 9 nearby core-collapse supernovae:
the 4 given in Table~\ref{table:recentsn} plus 2004am (3.5 Mpc), 2005af
(3.6 Mpc), 2002ap and 2003gd (both 7.3 Mpc), and 2002bu (about 7.5 Mpc).
The observed numbers of 9 within 10 Mpc (2.8 expected) and
4 within 4 Mpc (1.0 expected) indicate that the true nearby
supernova rates are probably about 3 times higher than in
our calculation, which we regard as quite conservative.

\begin{table}[t]
\caption{Selected nearby galaxies with high supernova rates.}
\label{table:recentsn}
\begin{ruledtabular}
\begin{tabular}{lcr}
Galaxy & D [Mpc] & Known Supernovae\\
\hline
NGC 2403 
& 3.3 
& 1954J, 2002kg, 2004dj \\
NGC 5236 (M 83)
& 4.5 
& 1923A, 1945B, 1950B, \\
& &1957D, 1968L, 1983N \\
NGC 6946 
& 5.9 
& 1917A, 1939C, 1948B, 1968D, \\
&&1969P, 1980K, 2002hh, 2004et \\
NGC 5457 (M 101)
& 7.4 
& 1909A, 1951H, 1970G \\
\end{tabular}
\end{ruledtabular}
\end{table}

\begin{figure}
\includegraphics[width=3.25in,clip=true]{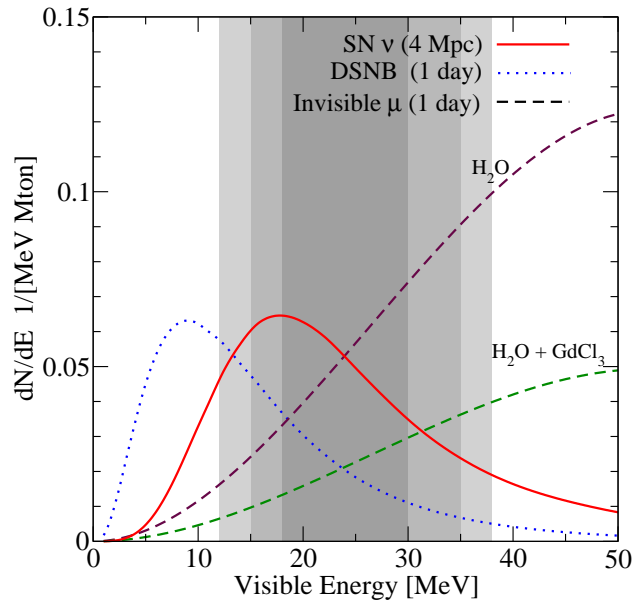}
\caption{Spectrum of detected supernova neutrinos ($D = 4$ Mpc) in a
1-Mton detector.  The backgrounds shown are relevant for a
neutrino-optical coincidence, and are reduced for a
neutrino-neutrino coincidence.  The shaded bands indicate the energy
intervals for signal selection (see Fig.~\ref{fig:prob}); at low energies, other
backgrounds (not shown) are very large.
\label{fig:spectra}}
\end{figure}

{\bf Neutrino-Neutrino Coincidence Detection.---}%
For a supernova in M 31, the yield in a 1-Mton detector would be high (about
50 events, over all energies).  However, the total nearby supernova rate remains
small until a distance of about 4 Mpc is reached, and then the number of detected
neutrinos per supernova is much smaller.  Thus we first consider the case in which
at least two candidate supernova neutrino events are detected within 10 seconds,
the supernova neutrino emission timescale. 
In Fig.~\ref{fig:spectra}, we show the expected neutrino signal in a
1-Mton detector for a supernova at 4 Mpc, using emission and
oscillation parameters similar to those in Ref.~\cite{Ando2002};
the 1-day backgrounds shown should be ignored here.   For
other reasonable choices of supernova neutrino temperatures and
oscillation scenario (i.e., an inverted hierarchy), the signal could
be significantly larger.   The detection
reaction is $\bar{\nu}_e + p \rightarrow e^+ + n$, for which the
visible positron energy is nearly the full neutrino energy~\cite{invbeta}.

The rate of accidental background coincidences within 10 s is small,
based on Super-Kamiokande data on spallation daughter decays~\cite{SKback1} and
invisible (sub-\v{C}erenkov) muon decays~\cite{SKback2}.  We estimate these
singles rates for a 1-Mton detector as $\alt 650$ yr$^{-1}$ above 15 MeV and 
$\alt 400$ yr$^{-1}$ below 35 MeV, respectively.  The total accidental coincidence
rate is thus $\alt 2\, (1050\ \mathrm{yr^{-1}})^2 (10 {\rm\ s}) = 0.7\ \mathrm{yr^{-1}}$,
scaling as the detector mass squared, and concentrated near the chosen energy
range boundaries ($15-35$ MeV).
The separation of signal and background events could easily be improved,
using the full energy and time distributions of events.
With at least two neutrinos detected, a supernova could be identified without
optical confirmation, so that the start of the light curve could be forecasted by a few
hours, along with a short list of probable host galaxies.  This would also allow the
detection of supernovae which are either heavily obscured by dust (e.g., in the
starburst galaxies M 82 and NGC 253) or are optically dark due to prompt black hole
formation.   If an optical supernova is found, even with crude
timing information, this would greatly reduce background rates.


{\bf Neutrino-Optical Coincidence Detection.---}%
To extend the reach to greater distances, we also consider the case in
which only one neutrino is detected, but a counterpart optical
supernova can be identified.  Although core-collapse supernova light
curves show a great deal of variation, we assume that it will be
possible to identify the start time of the core collapse to within
$\Delta t = 1$ day by optical techniques alone, at least for nearby
supernovae, which can be found very early.  In this case, only the
detector singles background rates are relevant, and these scale with
detector mass.  In Fig.~\ref{fig:spectra}, we show the spectrum
for the invisible muon background; nuclear gamma cuts are assumed in
the case of pure water, and also neutron cuts in the case of added
GdCl$_3$~\cite{GADZOOKS}.  The lower limits of the energy intervals
used in Fig.~\ref{fig:prob} are defined by large spallation and solar
neutrino backgrounds (pure water) and reactor backgrounds (with added
GdCl$_3$).

When an optical supernova is found, and its distance and start time
uncertainty $\Delta t$ identified, the neutrino data in the
appropriate energy interval can be checked.  Assuming that an optical
supernova is detected about once per year, the 1-day window
reduces the singles backgrounds by a factor 365.  Using
Fig.~\ref{fig:spectra}, one can estimate the probability that a
neutrino-like event within a 1-day interval of an optical
supernova was more likely signal or background.  In 18--30 MeV,
the numbers of signal and background events expected
are $N_\nu^{\mathrm{H_2O}} = 10 D_{\rm Mpc}^{-2} V_{\rm Mton}$ and
$N_{\rm bg}^{\mathrm{H_2O}} =  0.9 \Delta t_{\rm day} V_{\rm Mton}$;
in 12--38 MeV, they are $N_\nu^{\rm Gd} = 19 D_{\rm Mpc}^{-2} V_{\rm Mton}$
and $N_{\rm bg}^{\rm Gd} =  1.2 \Delta t_{\rm day} V_{\rm Mton}$. 
If one event is detected in association with an optical supernova, the
probability that it is real is $P_\nu / (P_\nu + P_{\rm bg})$, using the Poisson
probabilities for one event corresponding to $N_\nu$ and $N_{\rm bg}$.
At larger distances, the signal per supernova decreases as
$1/D^2$, but the supernova frequency increases as $D^3$.
Since the detector background rates are constant with $D$, the range
cannot be extended unless $\Delta t$ can be reduced, due to accidental
coincidences.


{\bf Discussion and Conclusions.---}%
We have proposed a new method for measuring the supernova neutrino
spectrum and for improving the observational characterization of nearby
core-collapse supernovae.  With a 1-Mton detector, supernova neutrinos
could be collected at a relatively brisk rate.  Considering just the galaxies
within 4 Mpc, and multiplying the supernova rate, the neutrino multiplicity,
and the neutrino detection probability (assuming added GdCl$_3$), we obtain
$0.3 \times 2 \times 0.25 \simeq 0.15$ and $0.3 \times 1 \times 0.4 \simeq 0.12$
neutrinos per year in the double and single detection modes, respectively. 
However, since the calculated supernova rates seem to be too conservative
by a factor of about 3, the total neutrino detection rates could be as large as
one per year.  The background rates are
comparable, but it should be possible to reduce their impact with a more
sophisticated analysis.

With the exception of SN 1987A in the Large Magellanic Cloud, a close
companion of the Milky Way, no neutrino source beyond the Sun has been
detected yet.  But besides the excitement of detecting neutrinos from
beyond 1 Mpc, and confirming core-collapse supernova neutrino emission,
there are quantitative reasons that detecting extragalactic supernova neutrinos
even one at a time would be important:

$\bullet$ {\it Measurement of the supernova neutrino spectrum:} The
$\simeq 20$ events from SN 1987A show statistically significant
disagreements with the predicted emission spectrum, and between
detectors~\cite{nu87A}.  A comparable number of events collected in new detectors
could resolve these issues and impact supernova $r$-process
nucleosynthesis calculations~\cite{rproc}.  These data would also
average over many supernovae, which is useful if the emission per
supernova is less uniform than expected, due to variation in the
properties and fates of the collapsed cores.  The $z = 0$ emission
spectrum could refine calculations of the redshift-integrated DSNB
flux, and the contribution of the harder nearby spectrum removed.
Since a typical distance is about 100 times greater than for SN 1987A,
limits on neutrino decay lifetimes could be correspondingly improved.
With each optical supernova providing the neutrino direction,
one could test the expected local galaxy clustering, the neutrino-positron
angular distribution~\cite{invbeta}, and possibly neutrino mixing effects in Earth.

$\bullet$ {\it A precise time trigger for other supernova signals:} We
have assumed that the start time of a nearby core-collapse supernova
can be determined to about 1 day using optical data alone.  This is
somewhat optimistic, though there is renewed interest in fully
characterizing the nearby core-collapse supernova rates and optical
emission.  For example, the Caltech Core-Collapse Project is designed
to extensively study 50 nearby core-collapse supernovae, in part to
explore the supernova--gamma-ray burst connection~\cite{CCCP}.  Early
supernova discoveries by amateurs may also be helpful.  The detection
of even a single neutrino in association with a nearby supernova would
reduce the uncertainty on the start time from $\sim 1$ day to $\sim
10$ seconds.  This precise trigger time could greatly reduce
backgrounds for more speculative types of prompt supernova emission,
e.g., gravitational waves~\cite{Arnaud,SNGW} and high-energy
neutrinos~\cite{RMW,SNjet}.


We thank D. DePoy, A. Gould, R. Pogge, and M. Vagins for discussions.
This work was supported by The Ohio State University.  S.~A. was
supported by a Grant-in-Aid for JSPS Fellows, and H.~Y. was supported
by NSF Grant No. PHY-0244384 at the University of Wisconsin, and by
the Wisconsin Alumni Research Foundation.


\end{document}